\documentclass[psfig]{aa}
\input psfig.tex

\begin{document}                                                                
\def\et{et al.}                                                                 
\def\egs{erg s$^{-1}$}                                                          
\def\egsc{erg s$^{-1}$ cm$^{-2}$}                                               
\def\msu{M$_{\odot}$\ }                                                         
\def\kms{km s$^{-1}$ }                                                          
\def\kmsM{km s$^{-1}$ Mpc$^{-1}$ }                                              
   \thesaurus{06         
              (03.11.1)}  

   \title{Likelihood Filter for Cluster Detection}            

   \author{P. Schuecker and H. B\"ohringer}
                                       
   \offprints{Peter Schuecker\\ peters@rosat.mpe-garching.mpg.de} 

   \institute{Max-Planck-Institut f\"ur extraterrestrische Physik, 
                   D-85748 Garching, Germany}

   \date{Received ..... ; accepted .....}                         
   
   \maketitle   
                                                                             
   \markboth {Likelihood Filter}{} 

\begin{abstract} The likelihood filter for cluster detection
introduced by Postman et al. is generalized by using standard
procedures and models originally developed in the theory of point
processes. It is shown that the filter formulae of Postman et al. can
be recovered in cases where background fields dominate the number
counts. The performance of the generalized method is illustrated by
using Monte Carlo simulations and by analyzing galaxy distributions
extracted from the COSMOS galaxy catalogue. The generalized method has
the advantage of being less biased at the expense of some higher
computational effort.

\end{abstract}

\section{Introducing Remarks}\label{S_IR}

The analogy between filtering and object detection in point
distributions was recognized since more than 25 years (e.g., Yashin
1970, Snyder 1972, Rubin 1972). Recent astronomical applications are
given in, e.g., Dalton et al. (1994), Kawasaki et al. (1998), and
Postman et al. (1996). In the latter project (Palomar Distant Cluster
Survey, PDCS) clusters of galaxies are detected and their redshifts
and richnesses are estimated by maximizing the excess of the galaxy
number counts as a function of angular coordinate and apparent
magnitude with respect to the foreground and background galaxy
distribution. In the following both distributions are referred to as
`background' distribution. The algorithm is based on the assumption
that the statistics are dominated by the background galaxy counts. In
this `high-background approximation' the expected large numbers of
galaxies ensure the validity of a Gaussian approximation of the
corresponding likelihood function (central limit theorem) which is
maximized to find the clusters and their basic parameters. It can be
shown that this type of maximization corresponds to the minimization
of the error-weighted squared differences between observed and modeled
data (method of least squares, e.g., Barlow 1996, p.\,93). All
counting errors are attributed to Poisson noise in the background. The
method outlined above is optimized in finding distant clusters where
the approximations seem to be appropriate. In the following we
describe a generalization of the PDCS filter where no {\it a priori}
information about the contrast of the cluster distribution with
respect to the background distribution is needed. We replace the
Gaussian approximation of the likelihood function by an exact
expression based on local Poisson distributions with
position-dependent mean intensities guided by the projected cluster
number density profile and by the luminosity function. A similar
approach based on binned data was developed independently by Kawasaki
et al. (1998). Their finally used likelihood function, however,
differs from the one derived in the present paper.

The aim of the present investigation is to study a mathematically
exact and general likelihood filter which can be further optimized and
then applied to the detection and characterization of isolated galaxy
clusters located in comparatively uniform background galaxy fields. We
plan to apply the method to galaxy fields centered around flux-limited
samples of X-ray sources, e.g., to the ROSAT-ESO-Flux-Limited X-Ray
(REFLEX) Cluster Catalogue (B\"ohringer et al., in preparation). The
algorithm can be applied, however, to a much broader class of
detection problems. The REFLEX clusters have measured redshifts up to
$z=0.30$; a few exceptional cases have $0.30<z<0.50$. The bulk of the
data is located in the range $0.02<z<0.15$.  Within the REFLEX
project, the filter can be used to compute statistical significances
of cluster detections, and to estimate redshifts and richnesses of the
optical counterparts. The redshift estimates could further support
those cluster redshifts which are obtained spectroscopically with only
small numbers of cluster galaxies. The richness estimates could give
more information about selection effects introduced by the complex
X-ray/optical survey process of the REFLEX project.

The approach is based on the assumption that the spatial and the
magnitude distributions of the galaxies located in the direction of
clusters can be modeled by a marked inhomogeneous Poisson point
process. The resulting likelihood function is given in a compressed
analytic form using the concept of the likelihood ratio statistics
(Sect.\,\ref{S_GM}). It is shown that the PDCS filter can be recovered
in cases where background number counts dominate the statistic
(Appendix\,B). Some practical notes on the application of the filter
are given in Sec.\,\ref{S_PC}. In Sect.\,\,\ref{S_FP} the performance
of the general method is illustrated by analyzing simulated data to
measure possible statistical biases of the derived numerical
estimators (very often maximum likelihood estimators are not unbiased;
see, e.g., Ripley 1991, Sect.\,4, and Barlow 1996, p.\,84) and by
analyzing subsamples of the COSMOS galaxy catalogue to test the
methods under more realistic survey conditions. The results are
summarized and the deviations from the idealized model assumptions are
discussed in Sect.\,\,\ref{S_CR}. All computations assume pressureless
Friedmann-Le\-ma\^{\i}\-tre world models with the Hubble constant,
$H_0$, in units of $h=H_0/(100\,{\rm km}\,{\rm s}^{-1}\,{\rm
Mpc}^{-1})$, a negligible cosmological constant, and the deceleration
parameter $q_0=0.5$.

\section{General Model}\label{S_GM}                                                    
The spatial distribution of the galaxies can be regarded as an
inhomogeneous but in principle more or less random pattern of points,
i.e., as a realization of a point process in mathematical terminology
(Neyman \& Scott 1952, Layzer 1956, see also Neyman 1961 and
references therein). The point process can be considered either as a
random set of discrete points or as a random measure, counting the
number of points in a given region (method of counts-in-cells). The
points may be distributed in a three-di\-men\-sion\-al volume, in a
two-di\-men\-sion\-al patch on the celestial sky, in a magnitude
space, etc.

\subsection {Inhomogeneous Poisson Point Process}\label{SS_IPPP}

All distributions with random locations of points and with local
density parameters guided by a {\it nonrandom} variable fall into the
category of inhomogeneous Poisson point processes. This is the case
for the inhomogeneous galaxy distribution seen in the direction of one
galaxy cluster where the local more or less radial-symmetric variation
of the projected galaxy number density profile plus the uniform
background field are determined by the intensity $\lambda(x)$. The
inhomogeneous Poisson point process is well-known in the theory of
point processes (see, e.g., Cressie 1993, p.\,650 and references
therein). If we replace the deterministic variation of the density
parameter by a {\it random} variable we enter the field of doubly
stochastic Poisson processes (Cox processes, see, e.g., Stoyan,
Kendall, \& Mecke 1995, p.\,154) which are the general types of point
processes to characterize large-scale distributions of galaxies (e.g.,
a Gaussian random field combined with local Poisson point processes).

Let ${x_1,\ldots,x_N}$ be one spatial realization of a point
process. The $N$ points, all with different coordinates, are
distributed within the total volume $A$. A useful quantity with a
simple interpretation and a direct relation to powerful analytical
tools provided by the theory of point processes is the local Janossy
density, $j_N(x_1,\ldots,x_N|A)$: after multiplication of $j(\cdot)$
with the product of the volume elements $dx_1\cdots dx_N$ it gives the
probability that in the total volume $A$ there are exactly $N$ points
in the process, one point in each of the $N$ distinct infinitesimal
half-open regions $[x_i,x_i+dx_i)$. The Janossy density may be
regarded as the likelihood $L_A$ of the realization, ignoring -- as
usual -- the principle differences between probability density
functions and sample functions:
\begin{equation}\label{LA}
L_A(x_1,\ldots,x_N)\,:=\,j_N(x_1,\ldots,x_N|A)\,.
\end{equation}
Assume that the number of galaxies observed per unit solid angle
depends only on the spatial coordinate $x$ and gives the intensity
$\lambda(x)$. More complicated models where also the magnitudes of the
galaxies are taken into account are discussed below in the context of
marked point processes (see Sec.\,\ref{SS_IMPP}). The intensity
$\lambda(x)$ is considered as the $x$-dependent density parameter of
an inhomogeneous Poisson process.

The Janossy density and thus the likelihood function of the
inhomogeneous Poisson point process can be derived directly by using
the machinery of probability generating functionals and Khinchin
measures (Daley \& Vere-Jones 1988, p.\,498). The application of this
formalism offers, however, the possibility to derive likelihood
functions for more complex point processes, e.g., multiple stochastic
point processes which might be interested for many cosmological
applications. A guideline through the basic equations of the formalism
is given in Appendix\,A. For inhomogeneous Poisson point processes
this standard formalism leads to the logarithmic likelihood (see also
Snyder 1975, Karr 1986, and Appendix\,A)
\begin{equation}\label{L1}
\ln L_A\,=\,-\int_A\lambda(x)dx\,+\,\sum_{i=1}^N\,\ln\lambda(x_i)\,.
\end{equation}
Equation (\ref{L1}) can also be found in a more heuristic way. As
already mentioned above, the quantity $L_A(x_1,$ $\ldots,x_N)$
$dx_1\cdots dx_N$ may be regarded as the probability of finding
exactly one point in each of the infinitesimal volume elements
$dx_1,\ldots,dx_N$, (case a), whereas outside of these regions no
further points are present, (case b). These events are independent,
due to the independence properties of the Poisson process. The
probabilities for (a) are given by $\lambda(x_i)dx_i$ (for each
individual point), and the probability for (b) is given by the
avoidance or void probability function, $\exp(-\int\lambda(x)dx)$. The
integration extents over the volume $A$ reduced by the sum of the
infinitesimal volume elements, $dx_i$, which can be neglected due to
their small size so that, in total, the integration extents over the
complete volume $A$ (Stoyan \& Stoyan 1992, p.\,258).

\subsection{Inhomogeneous Marked Poisson Point Process}\label{SS_IMPP}
           
In order to generalize the model described above note that galaxy
distributions can also be regarded as realizations of {\it marked}
spatial point processes, each consisting of locations of events in a
bounded study region $A$ and associated measurements (marks). Typical
marks, often found in cosmological applications are apparent
magnitude, luminosity, mass, energy, color, mor\-pho\-lo\-gi\-cal
type, etc. In this case the realization of a marked point process is
$(x_1,m_1),\ldots,(x_N,m_N)$. We assume the absence of any segregation
effects, i.e., that the marks are
independent-and-identically-distributed and are independent of the
associated marginal spatial point process. Define the point process by
a counting measure on the product space $A\times B$ of two intervals
with, for example, $x_i\in A\subset {\rm R}^3$, and $m_i\in B\subset
{\rm R}^1$. The moment measures of a marked spatial point process are
simple extensions of the moment measures of an ordinary spatial point
process. The generalization of (\ref{L1}) to marked inhomogeneous
Poisson point processes is thus straight\-forward and is given by
\begin{eqnarray}\label{LM1}
\ln L_{A\times B}\,&=&\,-\int_A\int_B\,\lambda(x,m) dm\,dx\,+\,
\sum_{i=1}^N\,\ln\lambda(x_i,m_i)\nonumber\\
 & &
\end{eqnarray}
Usually, $\lambda(x,m)$ can be factorized as
$\lambda(x,m)=f(m|x)\cdot$ $\lambda_0(x)$ where $f(m|x)$ is the
conditional density of the marks, so that the log-likelihood can be
maximized separately for the parameters characterizing $f$ and
$\lambda_0$.

\subsection{Application to Cluster Detection}\label{SS_ATCD}

In the case of finding clusters of galaxies using spatial and
magnitude information $\lambda(x,m)$ cannot be factorized as can be
seen from the model for the number of galaxies observed per unit area
and per unit magnitude interval introduced by Postman et al. (1996),
\begin{equation}\label{MUX}
\lambda(r,m)\,:=\,b(m)\,+\,\Lambda\,\phi(m-m^*)\,P\left(\frac{r}{r_c}\right)\,.
\end{equation}
Here, $b(m)$ gives the differential magnitude number counts of the
background galaxies, $\phi(m-m^*)$ is the luminosity function (e.g.,
in the Schechter prescription) of one cluster shifted along the
apparent magnitude scale in accordance with the given redshift
parameter and superposed onto the background distribution, $P(r/r_c)$
is the cluster angular surface number density profile as a function of
the projected (radial) distance $r$ from the center of the cluster
with the projected characteristic radius $r_c$, and $\Lambda$ is a
dimensionless parameter characterizing the intrinsic cluster
richness. Equation (\ref{MUX}) assumes a uniform, i.e., unclustered
background galaxy distribution, and that all clusters have the same
luminosity function and the same spatial number density profile. For
the model (\ref{MUX}) with given $b(m)$, and the normalization (if the
integrals do not diverge)
\begin{equation}\label{NORM}
\int_A\int_B\,\phi(m-m^*)\,P\left(\frac{r}{r_c}\right)\,dm\,d^2r\,=\,1\,,
\end{equation}
equation (\ref{LM1}) can be written as
\begin{eqnarray}\label{L2}
\ln L\,&=&\,-\,\int_A\int_B\,b(m)\,dm\,d^2r\,-\,\Lambda\,+\nonumber \\
&&\sum_{i=1}^N\,
\ln\left(b(m_i)+\Lambda\phi(m_i-m^*)P\left(\frac{r_i}{r_c}\right)\right)\,.
\end{eqnarray}
It is convenient to normalize likelihood functions using the concept
of likelihood ratio statistics. Further notes on the scope of
adaptability of the likelihood ratio statistics, especially when not
all random variables may be independent, can be found in Sen
\& Singer (1993, p.\,72). For the present situation where $b(m)$
is assumed to be known the results do, however, not explicitly depend
on the specific normalization. As a reference process in the
denominator usually the homogeneous Poisson process is chosen. In
order to get a likelihood ratio where its maximization explicitly
maximizes the contrast between cluster and background, an
inhomogeneous Poisson process with an $r$-independent density as given
in (\ref{MUX}) for $\Lambda=0$ (background model) seems to be more
appropriate. Note that although the point distribution of this
specific model is homogeneous in the spatial domain it is still
inhomogeneous in the magnitude space. The corresponding log-likelihood
is
\begin{equation}\label{L3}
\ln L_0\,=\,-\int_A\int_B\,b(m)\,dm\,d^2r\,+\,\sum_{i=1}^N\,\ln b(m_i)\,.
\end{equation}
With this normalization the final log-likelihood ratio is
\begin{eqnarray}\label{LR1}
\ln\left(\frac{L}{L_0}\right)\,=\,-\Lambda\,+\,\sum_{i=1}^N\,\ln\left(\,1\,+\,
\frac{\Lambda\phi(m_i-m^*)P(\frac{r_i}{r_c})}{b(m_i)}\,
\right)\nonumber\,.\\
\end{eqnarray}
The richness parameter $\Lambda$ is obtained from the relation
$\partial\ln(L/L_0)/\partial\Lambda=0$, resulting in the condition:
\begin{equation}\label{LAM1}
\sum_{i=1}^N\,\frac{\phi(m_i-m^*)\,P\left(\frac{r_i}{r_c}\right)}
{b(m_i)\,+\,\Lambda\,\phi(m_i-m^*)\,P\left(\frac{r_i}{r_c}\right)}\,=\,1\,.
\end{equation}
Equations (\ref{LR1}) and (\ref{LAM1}) can be used for the detection
of clusters in the following way: In the first step, values for $m^*$
and $r_c$ are selected so that, for given functional forms of $\phi$
and $P$, equation (\ref{LAM1}) gives the corresponding richness
parameter using standard numerical root finding algorithms (see, e.g.,
Brent 1973, Sect.\,3 and 4). In the second step, this $\Lambda$ value
is inserted into (\ref{LR1}) where $\ln(L/L_0)$ serves as a
significance measure for the detected cluster candidate. The relations
between the observed and the distant-independent values of the filter
parameters are determined by the redshift $z$ and the chosen
cosmological model. The final cluster redshift and richness values are
thus fixed by the $z$ value with the highest log-likelihood
ratio. From $\Lambda=0$ it can be deduced that for a general galaxy
field, $\ln(L/L_0)=0$ is the mode of the distribution of the
logarithmic values of the likelihood ratios. The distribution is
asymmetric with a long tail to larger $\ln(L/L_0)$-values where the
rich clusters are expected. Approximations of the equations
(\ref{LR1}) and (\ref{LAM1}) are given in Appendix\,B.

\section{Practical Considerations}\label{S_PC}

For the application of the filter the intervals $A$ and $B$ in
(\ref{NORM}) must be specified. The following computations assume the
half-open $r$- and $m$-ranges
\begin{equation}\label{INTERV}
A\,=\,[0,r_{\rm co})\,,\quad B\,=\,[m_{\rm bright},m_{\rm faint})\,.
\end{equation}
In equations (\ref{INTERV}) $r_{\rm co}$ is the halo or cutoff radius
defined by $P(r/r_c)=0$ for $r\ge r_{\rm co}$, and $P(r/r_c)>0$ for
$r< r_{\rm co}$. The bright and the faint-end limit of the apparent
magnitudes of the sample galaxies are $m_{\rm bright}$ and $m_{\rm
faint}$, respectively. For radial-symmetric cluster profiles,
equations (\ref{INTERV}), (\ref{MUX}), and (\ref{NORM}) give a second
estimator for the cluster richness,
\begin{eqnarray}\label{ESTRICH}
\hat{\Lambda}\,&=&\,2\pi\,\int_0^{r_{\rm co}}r\int_{m_{\rm bright}}^{m_{\rm
faint}}\, \lambda(r,m)dm\,dr\\ \nonumber &-&\,2\pi\,\int_0^{r_{\rm co}}r\int_{m_{\rm
bright}}^{m_{\rm faint}}\,b(m)dm\,dr\,=\,N_0\,-\,N_{\rm bg}\,= \,N_{\rm
cl}\,.
\end{eqnarray}
With the specific choice (\ref{INTERV}), $\hat{\Lambda}$ is the
difference between the total number of galaxies, $N_0$, seen in the
direction of the cluster, and the number of background galaxies,
$N_{\rm bg}$, expected for the cluster area determined by $r_{\rm
co}$. Therefore, $\hat{\Lambda}$ is a statistical measure of the
number of cluster galaxies, $N_{\rm cl}$, detected in the magnitude
range $B$ above the background distribution. The estimator gives
richnesses which depend on cluster redshift. Note the difference
between $\Lambda$ defined as a parameter characterizing the intrinsic
cluster richness by equation (\ref{MUX}) without any constrains on the
intervals $A$ and $B$, and $\hat{\Lambda}$ defined as an estimator of
the cluster richness through equation (\ref{ESTRICH}) under the
constrains (\ref{INTERV}). Equation (\ref{ESTRICH}) not only shows
that estimates of cluster richnesses obtained with (\ref{LAM1}) or
(\ref{ESTRICH}) are biased.  The equation also offers a new way to
circumvent the more complicated estimation of $\Lambda$ by solving
condition (\ref{LAM1}). Richnesses obtained with (\ref{ESTRICH}) are,
however, not obtained with the maximum likelihood
principle. Therefore, the combination of $\hat{\Lambda}$ with
(\ref{LR1}) does not necessarily yield consistent maximum likelihood
values.

A redshift-independent and thus unbiased estimate, $\Lambda_0$, of the
intrinsic cluster richness may be obtained from the ratios
\begin{eqnarray}\label{ESTRICH1}
\frac{\Lambda}{\Lambda_0}\,&=&\,\frac{\int_{m_{\rm bright}}^{m_{\rm
faint}}\,\phi(m-m^*)dm}{\int_{m_{\rm bright}}^{m_{\rm
co}}\,\phi(m-m^*)dm}\nonumber\\
&=&\,\left[\,\int_{m_{\rm bright}}^{m_{\rm co}}\,\phi(m-m^*)dm\,\right]^{-1}\,,
\end{eqnarray}
where $\Lambda$ is given by (\ref{LAM1}), and the cutoff magnitude,
$m_{\rm co}$, used as the upper limit of the numerical integration of
the luminosity function, must be chosen in accordance with the cluster
redshift. This magnitude may be defined by
\begin{equation}\label{INCR}
m_{\rm co}(z)\,:=\,m^*(z)\,+\,\Delta m\,\,[{\rm mag}]\,.
\end{equation}
Here, $m^*(z)$ is determined by the cluster redshift and $m_{\rm
co}(z)$ by a predefined increment, $\Delta m$ magnitudes fainter than
$m^*(z)$. The last equality in (\ref{ESTRICH1}) uses the normalization
(\ref{NORM}) separately for the radial and for the magnitude profile:
\begin{eqnarray}\label{NORM1}
\int_0^{r_{\rm co}}P\left(\frac{r}{r_c}\right)\,2\,\pi\,r\,dr=1\,,\quad
\int_{m_{\rm bright}}^{m_{\rm faint}}\phi(m-m^*)\,\,dm=1\,.\nonumber
\end{eqnarray}
The equations (\ref{ESTRICH1}) can also be used to correct
$\hat{\Lambda}$ for redshift biases. Finally, it should be noted that
$m_{\rm bright}$ must be bright enough to include the brightest galaxy
of the sample.

\section{Filter Performance}\label{S_FP}

The following computations illustrate the basic properties of the
likelihood filter. The model is given by the equations (\ref{LR1}),
(\ref{LAM1}), (\ref{INTERV}), (\ref{ESTRICH1}), and (\ref{INCR}). All
numerical simulations and reductions are performed for the
photographic $b_J$ passband. Observed data are taken from the COSMOS
galaxy catalogue (e.g, Heydon-Dumbleton et al. 1989).

For the magnitude filter, $\phi$, a Schechter-type luminosity function
is used with the characteristic magnitude, $M^*=-20.1+5\log h$\,[mag],
and the faint-end slope, $\alpha=-1.2$ (Colless 1989, Lumsden et
al. 1997). The COSMOS magnitudes are corrected for the effect of the
limited dynamic range within the microdensitometer (`saturation
effect') as in Lumsden et al. (1997, Sect.\,2.2). The cosmic $(K+E)$
corrections for the $b_J$ passband are parameterized as
$K(z)=3.0z$\,[mag] (Efstathiou, Ellis, \& Peterson 1988; Dalton et
al. 1997). If the galaxy magnitudes have large random errors one has
to replace $\phi(m-m^*)$ by the luminosity function convolved with the
magnitude errors.

The corrections for galactic extinction are applied to observed data
using the correlation between visual extinction and neutral Hydrogen
column density, $N_H[{\rm cm}^{-2}]/A_V$ $=1.79\times 10^{21}$ (Predehl
\& Schmitt 1995). The visual extinctions are transformed to the $B$
band using the colour excess $E_{B-V}=A_B-A_V$ and $A_B=4.3E_{B-V}$
(de Vaucouleurs et al. 1991, p.\,30). It is assumed that the
extinction in the $B$ band closely resembles the effect in the $b_J$
passband. Hydrogen column densities are taken from Dickey \& Lockman
(1992) as provided by the EXSAS97 release.

For the radial filter, $P$, a King-like profile as given in Postman et
al. (1996) is used,\\ 
$P(b)=\left\{\begin{array}{r@{\quad:\quad}l}
\left(1+b^2\right)^{-\frac{1}{2}}-
\left(1+C^2\right)^{-\frac{1}{2}} &
b<r/r_{\rm co}\\ 0 & b \ge r/r_{\rm co}
\end{array} \right.$\,,\\
with $b=r/r_c$ and $C=r_{\rm co}/r_c$. The core radius and the cutoff
radius have the respective values $r_c=100h^{-1}\,{\rm kpc}$ and
$r_{\rm co}=1h^{-1}$ ${\rm Mpc}$. For the richness estimates using
equation (\ref{ESTRICH1}) the magnitude increment $\Delta m =
3.0$\,mag in (\ref{INCR}) was chosen.

In the following paragraphs the detection of individual clusters and
the estimation of cluster richnesses are discussed
(Sect.\,\ref{SS_CRACDFS}).  For {\it rich} clusters the likelihood
filter derived in Sections\,\ref{S_GM} and \ref{S_PC} gives unbiased
estimates of the cluster redshift, as illustrated by the application
of the filter to simulated data (Sect.\,\ref{SS_NTFRB}) and to
observed data (Sect.\,\ref{SS_RFOD}).

\subsection{Cluster Richness and Cluster 
Detection from Simulations}\label{SS_CRACDFS}

\begin{figure}
\psfig{figure=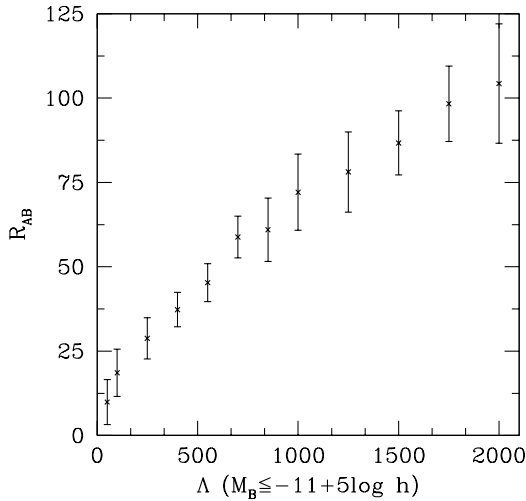,height=8cm}
\caption{{\small Abell cluster richness $R_{\rm AB}$ determined 
from simulations as a function of the intrinsic cluster richness
$\Lambda$, i.e., the number of cluster galaxies (above background)
with (as an example) absolute magnitudes $M\le
-11+5\log(h)$\,[mag]. The $1\sigma$ error bars are the standard
deviations obtained from simulations of clusters at different
redshifts.  The simulations give galaxy distributions as expected from
the COSMOS galaxy catalogue.}}\label{F_RICH}
\end{figure}

\begin{figure}
\psfig{figure=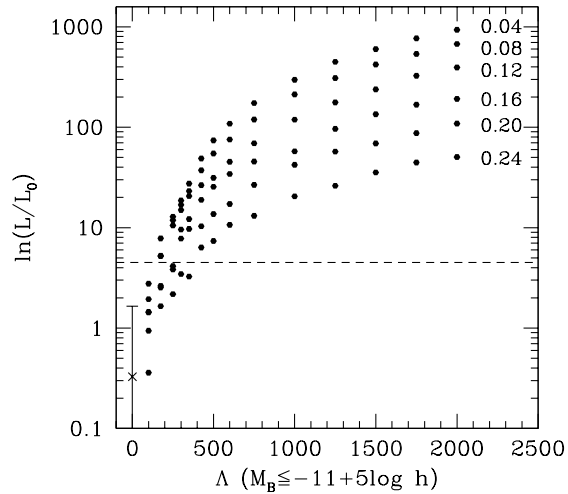,height=8cm}	
\caption{{\small Log-likelihood ratios for clusters (points) simulated with
intrinsic richnesses $100\le \Lambda \le 2000$ in the redshift range
$0.04\le z \le 0.24$, marked on the right-end of each sequence. The
short-dashed line shows the (formal) $3\sigma$ detection limit. The
vertical error bar at zero richness indicates the $3\sigma$ range of
the $\ln(L/L_0)$ values obtained for simulated fields without
superposed clusters of galaxies.}}\label{F_CDT1}

\end{figure}
\begin{figure}
\psfig{figure=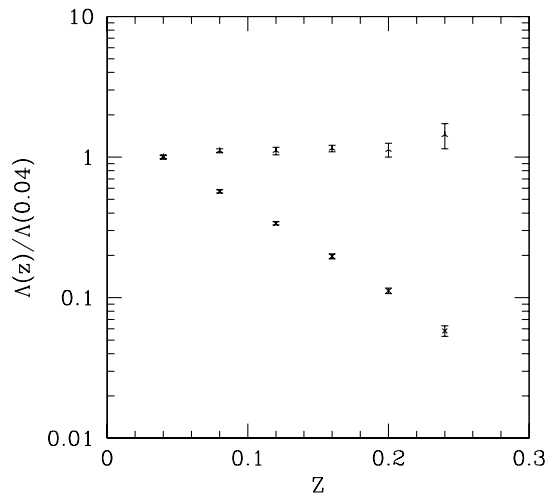,height=8cm}	
\caption{{\small Richness estimates $\Lambda$ (declining sequence)
and $\Lambda_0$ (horizontal sequence) as a function of redshift $z$
from simulated clusters of galaxies. Richnesses $\Lambda$ are computed
with equation\,(\ref{LAM1}); richnesses $\Lambda_0$ are computed with
equation\,(\ref{ESTRICH1}). Both richness estimates are normalized at
$z=0.04$. The error bars correspond to the $1\sigma$ standard
deviations. The errors are comparatively small because of the high
richness ($\Lambda=2000$) used in the simulations.}}\label{F_LAMBDA0}
\end{figure}

\begin{figure}
\psfig{figure=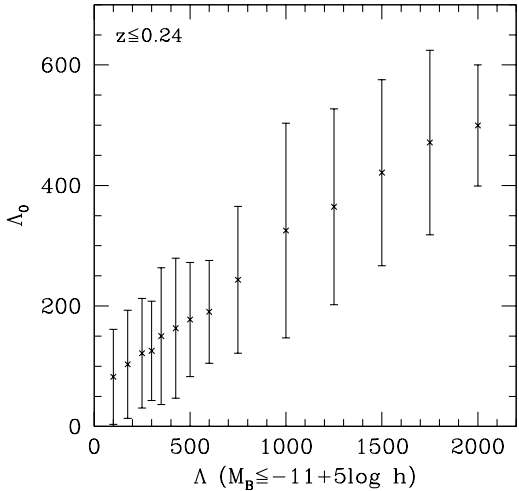,height=8cm}	
\caption{{\small Richness estimates $\Lambda_0$ obtained with the 
likelihood filter from simulated clusters with redshifts $z\le 0.24$
as a function of intrinsic richness $\Lambda$ (defined by
equation\,\ref{MUX}). Error bars correspond to $1\sigma$ standard
deviations.}}\label{F_LAMBDA1}
\end{figure}

\begin{figure}
\psfig{figure=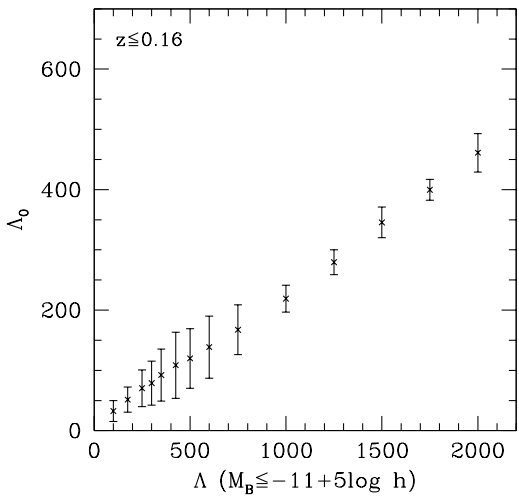,height=8cm}	
\caption{{\small As Fig.\,\ref{F_LAMBDA1} for clusters with 
redshifts $z\le 0.16$.}}\label{F_LAMBDA2}
\end{figure}

Although not essential for the application of the likelihood filter,
knowledge of the relation between the traditional Abell cluster
richness, $R_{\rm AB}$, and the not observable intrinsic cluster
richness, $\Lambda$, defined by equation (\ref{MUX}) is helpful in
order to understand the efficiency of cluster detection and cluster
characterization. Isolated clusters with known redshifts, equatorial
coordinates, and richnesses, superposed onto an homogeneous background
galaxy distribution, are simulated with simple Monte Carlo methods in
relativistic spacetime. The galaxies have apparent magnitudes in the
range $0.0\le b_J\le 20.5$\,mag. The background number counts, $b(m)$,
are guided by polynomial fits to the differential distributions
collected by Metcalfe et al. (1993). The luminosity functions and the
projected radial number density profiles of the simulated clusters are
the same as those used in the likelihood filter. The results obtained
from numerical simulations are shown in Fig.\,\ref{F_RICH}. For each
intrinsic richness $\Lambda$, the mean Abell richnesses (crosses) and
the $1\sigma$ error bars are obtained from clusters simulated at
different redshifts. The analysis concentrates on the Abell richness
classes $-1$,0,1, and 2 because most of the clusters we want to study
have these comparatively low richnesses. The nonlinearity between the
$R_{\rm AB}$ and the $\Lambda$ richnesses as seen in
Fig.\,\ref{F_RICH} is introduced by the richness-dependency of $m_3$,
that is, the apparent magnitude of the 3rd-brightest cluster galaxy
(Scott effect). The large scatter of the individual richness estimates
is mainly caused by the large scatter of the $m_3$ values.

Figure\,\ref{F_CDT1} shows the efficiency of the likelihood filter for
cluster detection as a function of the intrinsic cluster richness
$\Lambda$ and for different cluster redshifts, given at the right-end
of each sequence in the figure. The $3\sigma$ range of the
$\ln(L/L_0)$ values obtained for pure background fields without
superposed galaxy clusters is indicated by the vertical error bar at
$\Lambda=0$. As expected, the log-likelihood values increase with
increasing cluster richness and decreasing cluster redshift. The
simulations suggest that if the clusters have the same luminosity
function and radial number density profile as used in the likelihood
filter then the method would detect all clusters in the COSMOS galaxy
catalogue with redshifts $z\le 0.24$ and richness $\Lambda\ge 500$ or
$R_{\rm AB}>40$ (Abell richness class 0) on significance levels $\ge
3\sigma$. For larger redshifts, however, only clusters with larger
richnesses can be detected.

The richness estimates $\Lambda$ obtained with equation\,(\ref{LAM1})
and the richness estimates $\Lambda_0$ obtained with (\ref{ESTRICH1})
are compared in Fig.\,\ref{F_LAMBDA0}. As shown in Sect.\,\ref{S_PC},
$\Lambda$ estimates are always redshift-de\-pen\-dent when they are
computed within fixed magnitude intervals. They give the number of
cluster galaxies above the background field: this number decreases
with increasing $z$. Contrary to this, $\Lambda_0$ estimates are
corrected for the effect of the redshift-dependent magnitude range of
the cluster galaxies. These estimates are almost constant over the
given $z$ interval.

The accuracies of richness estimates are always limited by principle
problems concerning the determination of the cluster memberships and
by the small numbers of detectable galaxies in distant clusters. These
problems are illustrated in the Figs.\,\ref{F_LAMBDA1} and
\ref{F_LAMBDA2}. The richness estimates $\Lambda_0$ are computed for
clusters at different redshifts and intrinsic cluster richnesses
$\Lambda$. The number of simulated clusters is constant per unit
redshift interval. As expected, the random errors of the $\Lambda_0$
estimates increase significantly for poor clusters at high
$z$. Moreover, cluster richnesses are systematically overestimated for
samples with high $z$. Only for cluster samples with small $z$ and
high $\Lambda$ we can expect unbiased estimates which are directly
linked to the intrinsic cluster richnesses.

\subsection{Tests for Redshift Biases from Simulations}\label{SS_NTFRB}

The following tests are devoted to studies concerning possible
systematic redshift errors which might be inherent for the derived
method. In order to detect $z$ biases without the disturbing effects
of the discreteness noise, the analysis begins with the simulations of
clusters with high intrinsic richnesses, $\Lambda=10^4$. We regard
these simulations as some kind of numerical integration of the
relevant equations. At the end of this paragraph, the results from the
simulations of clusters with low richnesses are discussed.

\begin{figure}
\psfig{figure=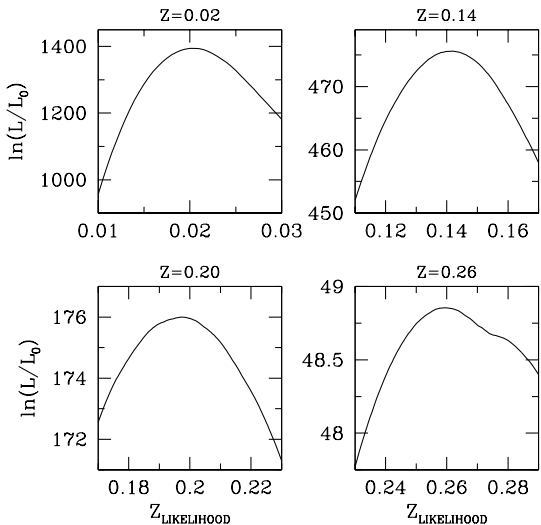,height=8cm}
\caption{{\small Examples of log-likelihood curves from simulated
data.  The intrinsic richness of the clusters is $\Lambda=10^4$. The
redshifts of the clusters are given on top of each panel. Formal
standard deviations of the redshift estimates can be obtained from the
points where the log-likelihood curve has fallen by 0.5 ($1\sigma$),
2.0 ($2\sigma$), 4.5 ($3\sigma$), $\ldots$} from its
maximum.}\label{F_LHC}
\end{figure}

\begin{figure}
\psfig{figure=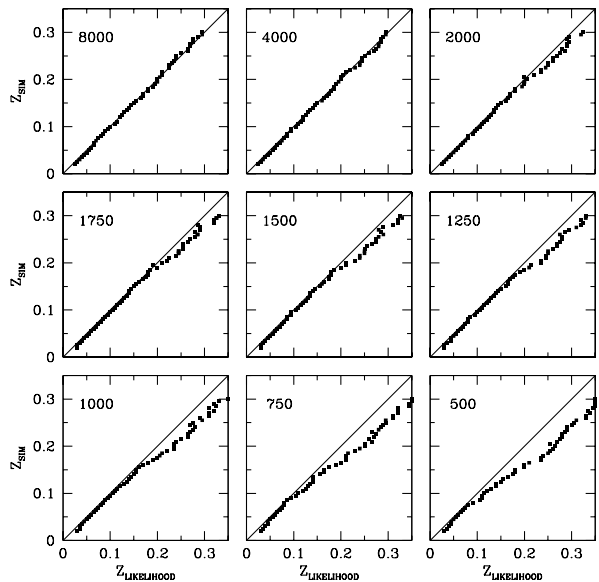,height=8cm}	
\caption{{\small Comparison of the redshift estimates obtained with the
likelihood filter, $z_{\rm Likelihood}$, with the true cluster
redshifts, $z_{\rm SIM}$, for different intrinsic cluster richnesses
($\Lambda$ value given in the upper left of each panel). The data are
obtained from simulations. The richness $\Lambda=8000$ corresponds to
a very rich cluster whereas $\Lambda=500$ characterizes a cluster
which is near the detection limit of the likelihood filter expected
for the simulated COSMOS data when the cluster is located at $z\approx
0.24$.}}\label{F_ZLIKZSIM}
\end{figure}

Figures\,\ref{F_LHC} show the likelihood values obtained for four rich
clusters simulated at the redshifts $z=0.02$, 0.14, 0.20, and 0.26.
As already shown in the preceding sections, the magnitude and
redshift ranges of these simulations give galaxy distributions as
selected from the COSMOS galaxy catalogue.

For all redshifts the log-likelihood curves in Fig.\,\ref{F_LHC} peak
close to the true cluster redshifts (indicated on top of each
panel). For intermediate $z$ values, the comparatively high symmetry
of the curves around the correct cluster redshift indicates the
absence of large systematic $z$ errors. For small and for large
redshifts the curves get more asymmetric. This might be taken as a
hint to possible increasing systematic $z$ errors when the method is
applied to comparatively nearby or distant clusters.

The asymmetries, however, do not significantly affect the positions of
the maxima of the log-likelihood curves and thus of the derived
redshifts. For example, the differences, $\Delta z$, between the true
redshifts and the likelihood redshifts are $|\Delta z|=0.0005$ for the
extreme redshifts $z=0.02$ and $z=0.26$, and do not exceed $|\Delta
z|=0.0025$ for all cases studied so far. For very rich clusters, the
$\Delta z$ values do not show any $z$-dependent trends, and are of the
order of the expected numerical accuracies.

We thus conclude that numerical studies under ideal conditions (high
richness, and background distribution assumed to be known) give no
indications for possible $z$-dependent biases of the locations of the
maxima of the log-likelihood curves.

The situation changes for poor clusters and when the background number
counts must be estimated from the sample itself. The comparison of the
likelihood redshifts with the true cluster redshifts is shown for
different intrinsic richnesses in Fig.\,\ref{F_ZLIKZSIM}. For high
richnesses no systematic redshift errors are detected, supporting the
results of the analyses described above. For poor clusters with, e.g.,
$\Lambda=500$ (i.e., clusters close to the detection limit at
$z=0.24$) a systematic overestimation of $z$ at high redshifts is
found. This is mainly caused by the small number of cluster galaxies
detected for poor clusters at high $z$ leading to a systematic
undersampling of the bright end of the cluster luminosity function and
of the outskirts of the projected cluster profile.

\subsection{Redshifts from Observed Data}\label{SS_RFOD}

\begin{figure}
\vspace{-2.5cm}
\hspace{-0.0cm}
\psfig{figure=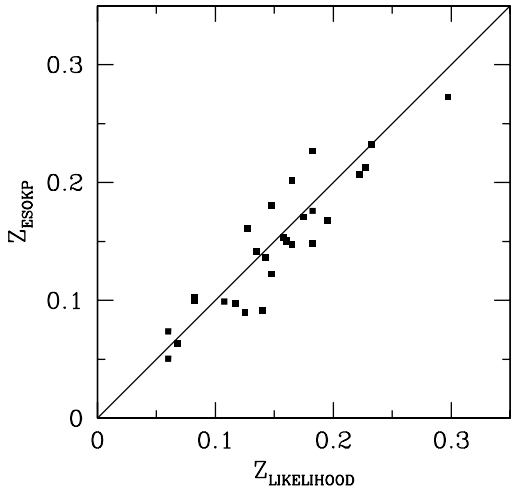,height=10cm}
\caption{{\small Redshifts of observed clusters of galaxies, 
$Z_{\rm LIKELIHOOD}$, obtained with the likelihood filter using
subsamples of the COSMOS galaxy catalogue, and spectroscopic cluster
redshifts, $Z_{\rm ESOKP}$, obtained with eight to 20 galaxies per
cluster as published in the REFLEX catalogue of B\"ohringer et al. (in
preparation).}}\label{F_ZCOM}
\end{figure}

Before we describe the results obtained with the likelihood filter on
observed data a few comments regarding some practical limitations are
given. The derivation of the likelihood filter rests on the assumption
of an homogeneous and isotropic spatial distribution of the background
galaxies. In mathematical terminology this involves the validity of
first-order motion invariance, i.e., the invariance of the mean
intensity of galaxies in the study area outside the cluster region
under spatial translations and rotations. In many practical situations
the observed data do not fulfill these requirements.

On small scales, the galaxy field is strongly clustered so that
comparatively large solid angles are needed to reduce biases in the
background determination. On large scales, extreme care must be taken
to achieve homogeneous sampling of the galaxy field. The latter aspect
is important for the analysis of galaxy catalogues obtained from
mosaics of digitized Schmidt plates where the effects of vignetting,
desensitization, and small errors in magnitude zero-points may
introduce artifical fluctuations in the observed galaxy field,
especially when galaxy distributions obtained from different plates
are combined. For the application of the likelihood filter to observed
data it is thus necessary to extract galaxy fields on areas which are
not too small to be affected by galaxy number density fluctuations and
which are not too large to be affected by artifical plate-to-plate
variations.

Tuckey's median polishing offers a robust way to detect large-scale
deviations from first-order motion invariance in point
distributions. The algorithm successively subtracts medians out of
rows of gridded data (number counts in quadratic cells), then columns,
then rows, then columns, and so on, accumulating them in `row',
`column', and `all' registers, and leaves behind the table of
residuals. The convergence of this procedure is discussed in Kemperman
(1984). An illustration of the algorithm is given in Appendix\,C.

We use this algorithm after transforming the equatorial object
coordinates into Ham\-mer-Aitoff equal-area coordinates to flag
regions by searching for cells in the row and in the column registers
which significantly deviate ($>5\sigma$ formal background noise) from
the mean galaxy background distribution and exclude the corresponding
plates from further analyses. The results obtained with median
polishing sometimes strongly depend on the specific orientation of the
row and column axes. Therefore, we apply the algorithm using three
different orientations. For each orientation, median polishing usually
converges after three to five iterations. The iteration stops when the
relative changes between two successive iteration steps are less than
one percent.

In Fig.\,\ref{F_ZCOM} the redshift estimates obtained with the
likelihood filter on subsamples of the COSMOS galaxy catalogue are
compared to cluster redshifts obtained from spectroscopic
observations. The central cluster coordinates and the cluster
redshifts are taken form the REFLEX catalogue (B\"ohringer et al., in
preparation). The test sample includes 43 REFLEX clusters observed
among others within the ESO Key Programme (e.g., Guzzo et
al. 1995). These clusters have redshifts determined with eight to 20
cluster galaxies and are thus well-suited as redshift references. In
addition, the presence of comparatively strong X-ray signals from
these sources and the absence of active galaxies further support the
existence of real clusters in these areas. The test sample is not a
representative subsample of the complete REFLEX catalogue. It is
biased to larger redshifts because the complete REFLEX sample includes
also a large number of nearby clusters.

All COSMOS galaxies with $b_J\le 20.5$\,mag (after correction for
saturation effects and galactic extinction) located within spherical
regions of about two degrees radius around the X-ray cluster
positions are selected. After median polishing and careful
determination of the bright and of the faint magnitude limits of the
galaxies in each study region, typically 10\,000 galaxies (background
$+$ cluster) remain per cluster target.

For 28 clusters from the list of 43 test clusters the filter detects a
maximum in the likelihood functions within the redshift range $0.0\le
z \le 0.35$.  The parameters of the likelihood filter are the same as
used in the simulations. The corresponding $Z_{\rm LIKELIHOOD}$ values
are shown in Fig.\,\ref{F_ZCOM}. For the remaining 15 test clusters no
redshifts could be determined, either because of their weak
statistical significance $\ln(L/L_0)<4.5$ (two cases) or because no
local maximum was detected (13 cases). These redshift incompletenesses
increase with increasing redshift from 20 to 25 percent for $z<0.2$ to
53 percent for $z>0.2$. However, the redshift distribution of the test
clusters was constructed to be almost redshift-independent so that the
given incompleteness values must be weighted with the {\it a priori}
probabilities for the presence of clusters of galaxies within
different redshift intervals (multiplication formula for
probabilities). Therefore, the actual incompletenesses are expected to
be much smaller compared to the values given above. The standard
deviation of the redshift differences $(Z_{\rm LIKELIHOOD}-Z_{\rm
ESOKP})$ is $\sigma_z=0.023\pm 0.004$.

\section{Concluding Remarks}\label{S_CR}
\begin{table}
\caption{Effects of false $M^*$ and $r_{\rm co}$ filter parameters 
on redshift and richness estimates. $M^*$ values are given in units of
$5\log(h)$\,mag, $r_{\rm co}$ values are given in units of Mpc$/h$ (see
text).}
\label{T_MRCO}
\[
\begin{array}{rrrrrrrrrr}
\hline
\noalign{\smallskip}
M^* |& r_{\rm co} & 0.5   &      & &  1.0 &      & &  2.0 &  \\
\noalign{\smallskip}
\hline \\
\noalign{\smallskip}
-18.1&           & 0.050 & 1033 & &0.075 & 2080 & &0.090 & 3029\\
-20.1&           & 0.064 &  433 & &\fbox{0.109} &\fbox{986} & &0.148 & 1754\\
-22.1&           & 0.065 &  206 & &0.131 &  425 & &0.240 & 1090\\
\noalign{\smallskip}
\hline
\end{array}
\]
\end{table}

\begin{table}
\caption{Effects of false $M^*$ and $r_{\rm c}$ filter parameters 
on redshift and richness estimates. $M^*$ values are given in units of
$5\log(h)$\,mag, $r_{\rm c}$ values are given in units of Mpc$/h$ (see
text).}
\label{T_MRC}
\[
\begin{array}{rrrrrrrrrr}
\hline
\noalign{\smallskip}
M^* |& r_{\rm c} & 0.05  &      & &  0.1 &      & &  0.5 &  \\
\noalign{\smallskip}
\hline \\
\noalign{\smallskip}
-18.1&           & 0.070 & 1835 & &0.075 & 2080 & &0.085 & 2638\\
-20.1&           & 0.106 &  921 & &\fbox{0.109} &\fbox{986} & &0.120 & 1121\\
-22.1&           & 0.128 &  420 & &0.131 &  425 & &0.140 &  457\\
\noalign{\smallskip}
\hline
\end{array}
\]
\end{table}

\begin{table}
\caption{Effects of substructure on the richness and redshift
estimates as a function of the separation $\Delta r$ and for different
richness ratios $\Lambda_1/\Lambda_2$ of the two components. The
application of the general likelihood filter gives the normalized
richness (numerator) and the redshift (denominator) for the combined
system.}
\label{T_SUBSTR}
\[
\begin{array}{lrrrrrrr}
\hline
\noalign{\smallskip}
\frac{\Delta r}{r_c} \,|& \frac{\Lambda_1}{\Lambda_2} & 1.000 
& 0.500 & 0.250 & 0.125 & 0.025 & 0.000 \\
\noalign{\smallskip}
\hline \\
\noalign{\smallskip}
0 & & \frac{2.059}{0.104} & \frac{1.529}{0.105} &
\frac{1.265}{0.105} & \frac{1.130}{0.106} & \frac{1.026}{0.106} 
& \frac{1.000}{0.106} \\
\noalign{\smallskip}
1 & & \frac{2.049}{0.104} & \frac{1.525}{0.105} &
\frac{1.409}{0.105} & \frac{1.259}{0.106} & \frac{1.025}{0.106} 
& \frac{1.000}{0.106} \\
\noalign{\smallskip}
2 & & \frac{2.043}{0.101} & \frac{1.517}{0.104} &
\frac{1.255}{0.105} & \frac{1.257}{0.106} & \frac{1.025}{0.106} 
& \frac{1.000}{0.106} \\
\noalign{\smallskip}
3 & & \frac{2.023}{0.102} & \frac{1.505}{0.105} &
\frac{1.246}{0.104} & \frac{1.122}{0.105} & \frac{1.024}{0.105} 
& \frac{1.000}{0.106} \\
\noalign{\smallskip}
5 & & \frac{1.759}{0.095} & \frac{1.464}{0.103} &
\frac{1.229}{0.101} & \frac{1.110}{0.104} & \frac{1.022}{0.105} 
& \frac{1.000}{0.106} \\
\noalign{\smallskip}
7 & & \frac{1.511}{0.104} & \frac{1.254}{0.102} &
\frac{1.188}{0.096} & \frac{1.086}{0.101} & \frac{1.016}{0.106} 
& \frac{1.000}{0.106} \\
\noalign{\smallskip}
\hline
\end{array}
\]
\end{table}

Both the PDCS filter and the generalized filter derived in the present
paper give systematic redshift errors for clusters at high $z$. The
generalized filter overestimated redshifts at high $z$; the errors
tend to zero for richer systems. The PDCS underestimates redshifts at
high $z$; the errors tend to zero for richer systems (see Fig.\,14 in
Postman et al.). Whereas the errors of the generalized method can be
explained by the incomplete sampling of the filter profiles for poor
clusters at high $z$, the errors of the PDCS filter can not be judged
in an easy way. Reasons for the descrepancies between the two methods
are still not clear because of the complexity of the PDCS
approximations. Nevertheless, it seems that if one uses the exact
mathematical equations, as does the generalized method, corrections
like the `cluster signal correction' of the PDCS filter are not
necessary.

It it shown that the richness estimates obtained with the generalized
filter are redshift-dependent and must be corrected (e.g., with
equation\,\ref{ESTRICH1}).  It is not clear why similar corrections
are not necessary for the PDCS filter.

Similar conclusions about the PDCS filter are drawn by Kawasaki et
al. (1998). In contrast to the generalized method, the Kawasaki filter
uses binned data and does not maximize the contrast between field and
cluster galaxy counts. A detailed comparison of the generalized method
and the approximations introduced by Kawasaki et al. is in
preparation.

The final question concerns the validity of the chosen assumptions
used to derive the likelihood filters. Problems caused by deviations
of the spatial distribution of the background galaxies from the
assumed uniform distribution were already discussed in
Sect.\,\ref{SS_RFOD}.

Further systematic errors in the derived cluster redshifts and
richnesses arise when the luminosity functions and the number density
profiles of the likelihood filter do not match with the corresponding
functions of the observed clusters. As an example, Tabs.\,\ref{T_MRCO}
and \ref{T_MRC} illustrate the effects of false filter parameter
values $M^*$, $r_{\rm co}$, and $r_c$ on the estimates of $z$ and
$\Lambda_0$. The results are obtained from simulated data.  The
`correct' cluster parameters, derived for the case when $\phi$ and $P$
of the filter exactly match with the corresponding cluster functions,
are shown as reference in the tables within the small boxes.

The ranges of the filter parameters are clearly far beyond the
variations expected for realistic cases. Nevertheless, the results
given in Tabs.\,\ref{T_MRCO} and \ref{T_MRC} illustrate the
stabilizing effect of using both spatial and magnitude information to
derive the cluster $z$ and $\Lambda_0$. For some parameter
combinations, however, the systematic errors are amplified.

It is also seen that false values of the cutoff radius, $r_{\rm co}$,
give larger systematic $z$ and $\Lambda_0$ errors compared to false
values of the core radius, $r_c$. One might expect that the stronger
effect of $r_c$ on truncated cluster profiles would also translate
into higher sensitivities of $z$ and $\Lambda_0$ on this filter
parameter. The conclusion, however, neglects the effects of errors in
the background number counts. The number of the background galaxies
expected for the cluster area scales with $r^2_{\rm co}$ so that small
errors in $r_{\rm co}$ have a large effect.

In many clusters of galaxies the presence of substructures may lead to
significant systematic errors in the estimated cluster richnesses and
redshifts. In order to simulate unrelaxed clusters in a stage of
formation by merging of subunits we approximate such configurations by
superposing two cluster models at close separation which seems to
describe observed mergers quite well (e.g., Briel et al. 1991,
B\"ohringer et al. 1996). The two components are supposed to have the
same redshift and therefore the same number density profiles and
luminosity functions, but varying richness ratios are
chosen. Tab.\,\ref{T_SUBSTR} summaries the results obtained for
different distances $\Delta r$ (in units of $r_c$) between the centers
of the two components, and for different ratios,
$\Lambda_1/\Lambda_2$, of the richnesses. The table gives for the
combined system the estimated richnesses (numerator) normalized to the
richness of the main component, and the estimated redshifts
(denominator). As expected, the maximum likelihood filter gives for
small $\Delta r$ the approximate sum of the two richnesses, converging
for large $\Delta r$ to the richness of the main
component. Fortunately, the redshifts are almost uneffected by this
type of substructure.

The effects of different faint-end slopes of the luminosity functions
(not shown here) are significantly smaller compared to the effects
mentioned above and can in general be neglected for the distant
clusters.

To summarize, the likelihood filter described here, is shown to be a
useful method to detect and to characterize distant clusters of
galaxies. Clusters with comparatively low richnesses can be detected
out to fairly large distances. Reliable redshift and richness
estimates, can, however, only be obtained for the more nearby, richer,
and isolated systems.

\begin{acknowledgements}                                                        
Sincere thanks go to H.T. MacGillivray for making the COSMOS galaxy
catalogue available to us. We thank the referee, A. Bijaoui, for some
useful comments, especially for his idea to study the effects of
substructures. P.S. thanks for support by the Verbundforschung under
the grant No.\,50\,OR\,9708\,35.

\end{acknowledgements}

\begin{appendix}

\section{Derivation of Likelihood Functions for Point Processes}
\label{S_DOLFFPP}

In the first step, the probability generating functional $G_A$ of a
point processed, defined by the expectation
\begin{eqnarray}\label{GF}
G_A[h]\,&:=&\,{\rm E}\left(\,\prod_{x_i\in
{\rm N}}\,h(x_i)\right)\,,
\end{eqnarray}
must be specified (e.g., Stoyan, Kendall, \& Meck 1995, p.\,116). Here,
$h(x)$ is a $[0,1]$-valued (test) function. In (\ref{GF}) the point
process on $A$ is represented by ${\rm N}$, and the points are located
at the spatial positions $\{x_i\}$.

In the second step, the logarithmic generating functional obtained
with (\ref{GF}) is compared with the expansion
\begin{eqnarray}\label{GAK}
\ln G_A[h]\,&=&\,-K_0(A)\,+\,\sum_{n=1}^\infty\,(n!)^{-1}\,
\int_{A^{(n)}}\nonumber \\ & &\cdots\int\,h(x_1)\cdots h(x_n)\,K_n(dx_1\cdots dx_n|A)\,,
\end{eqnarray}
to get the Khinchin measures, $K_n$. Some remarks concerning the
derivations of both equation (\ref{GAK}) and (\ref{JN1}, see below)
can be found in Daley \& Vere-Jones (1988, Sect.\,5.5, and p.\,230).

In the last step, the $K_n$ measures are inserted into the exansion of
the local Janossy densities,
\begin{eqnarray}\label{JN1}
&&j_N(x_1,\ldots,x_N|A)\,=\nonumber \\&&e^{-K_0}
\,\sum_{r=1}^N\,\sum_{\tau\in{\cal P}_{rN}}\,
\prod_{i=1}^r\,k_{|S_i(\tau)|}(x_{i,1},\ldots,x_{i,|S_i(\tau)|}|A)\,\,,
\end{eqnarray}
to get in combination with equation\,(\ref{LA}) the likelihood
function of the point process. The second sum on the right-hand side
of (\ref{JN1}) is taken over all $r$-partitions $P_{rN}$ where
$|S_i(\tau)|$ gives the number of elements in each partition set.

Example: Inhomogeneous Poisson point process.\\ For this process,
equation (\ref{GF}) gives (Cressi 1993, p.\,650)
\begin{equation}\label{GA}
\ln G_A[h]\,=\,-\int_A\,(1-h(x))\,\lambda(x)\,dx\,.
\end{equation}
With equation (\ref{GAK}) we obtain the only nonzero Khinchin
measures, $K_0=\int_A\lambda(x)dx$ and $k_1(x|A)=\lambda(x)$,
resulting to the Janossy density
\begin{equation}\label{JN2}
j_N(x_1,\ldots,x_N|A)\,=\,e^{-K_0}\,\prod_{i=1}^N\,k_1(x_i|A)\,,
\end{equation}
which directly leads to the likelihood function of the inhomogeneous
Poisson point process (eq.\,\ref{L1} in Sect.\,\ref{SS_IPPP}).

\section{High-Background Approximation}\label{S_HBA}

Equations (\ref{LR1}) and (\ref{LAM1}) are mathematically exact and
are based on no special mathematical approximation. In this section it
will be shown that the equations of the digital filter derived by
Postman et al. (1996) -- not the finally equations used -- can be
obtained if
\begin{equation}\label{HBA}
b(m)\,\gg\,\Lambda\,\phi(m-m^*)\,P\left(\frac{r}{r_c}\right)\,.
\end{equation}
In this limit the left side of the condition (\ref{LAM1}) may be
approximated by
\begin{eqnarray}\label{HBA1}
\sum_{i=1}^N\,\frac{\phi(m_i-m^*)}{b(m_i)}P\left(\frac{r_i}{r_c}\right)\,-\,
\Lambda\sum_{i=1}^N
\frac{\phi^2(m_i-m^*)}{b^2(m_i)}P^2\left(\frac{r_i}{r_c}\right)\nonumber\,,
\end{eqnarray}
from which a closed analytic form for the cluster richness is obtained
\begin{equation}\label{HBA2}
\Lambda\,\approx\,
\frac{\sum_{i=1}^N\,\frac{\phi(m_i-m^*)}{b(m_i)}P(\frac{r_i}{r_c})\,-\,1}
{\sum_{i=1}^N\,\frac{\phi^2(m_i-m^*)}{b^2(m_i)}P^2(\frac{r_i}{r_c})}\,.
\end{equation}
In the same limit (\ref{LR1}) reduces to
\begin{equation}\label{HBA3}
\ln\left(\frac{L}{L_0}\right)\,\approx\,\Lambda\,\left(
\sum_{i=1}^N\,\frac{\phi(m_i-m^*)}{b(m_i)}
P\left(\frac{r_i}{r_c}\right)\,-\,1\right)\,.
\end{equation}
Inserting (\ref{HBA2}) into (\ref{HBA3}) gives an analog equation to
(\ref{LR1}) in the high-background approximation
\begin{equation}\label{HBA4}
\ln\left(\frac{L}{L_0}\right)\,\approx\,
\frac{\left(\sum_{i=1}^N\,
\frac{\phi(m_i-m^*)}{b(m_i)}P\left(\frac{r_i}{r_c}\right)\,-\,1\right)^2}
{\sum_{i=1}^N\,
\frac{\phi^2(m_i-m^*)}{b^2(m_i)}P^2\left(\frac{r_i}{r_c}\right)}\,.
\end{equation}
It is seen that equations (\ref{HBA4}) and (\ref{HBA2}) correspond to
equations (15) and (14) in Postman et al. (1996), respectively, if
\begin{eqnarray}\label{IF1}
&&\sum_{i=1}^N\,\frac{\phi^2(m_i-m^*)P^2\left(\frac{r_i}{r_c}\right)}
{b^2(m_i)}\nonumber \\ &\approx&\,
\int_A\int_B\,\frac{\phi^2(m-m^*)P^2\left(\frac{r}{r_c}\right)}{b^2(m)}
\lambda(r,m)\,dm\,d^2r
\end{eqnarray}
\begin{equation}\label{IF2}\approx\,
\int_A\int_B\,\frac{\phi^2(m-m^*)P^2\left(\frac{r}{r_c}\right)}{b(m)}\,dm\,d^2r\nonumber\,,
\end{equation}
and
\begin{eqnarray}\label{IF3}
&&\sum_{i=1}^N\,\frac{\phi(m_i-m^*)P\left(\frac{r_i}{r_c}\right)}
{b(m_i)}\nonumber\\ &\approx&\,
\int_A\int_B\,\frac{\phi(m-m^*)P\left(\frac{r}{r_c}\right)}{b(m)}\,\lambda(r,m)\,dm\,d^2r\,,
\end{eqnarray}
and the normalization (\ref{NORM}) holds.

\section {Median Polish Algorithm}\label{S_MPOL}

In the present investigation median polishing is used to identify
those photographic plates where the surface number density of galaxies
deviates significantly from the central plate containing the major
part of the galaxy cluster, so that the galaxies on those
`pathological' plates can be rejected from further analyses. The
algorithm is, however, more powerful and thus interesting enough to be
described in more detail. The method can, for example, be used for
samples of sufficiently large sizes as a {\it definition procedure} of
stationary point distributions.

Let $N_{kl}$, $k=1,\ldots,p; l=1,\ldots,q$ be the number of points
counted in the cell with the indices $k$ and $l$. The $k$ and the $l$
index may number the count cells along the Right Ascension and along
the Declination axis, respectively. We follow the terminology of
Cressi (1993, p.\,186) and define for $i=1,3,5,\ldots,$
\begin{eqnarray}\label{MPOL1}
N^{(i)}_{kl}\,&:=&\,N^{(i-1)}_{kl}\,-
\,{\rm med}\{N^{(i-1)}_{kl}\,|\,l=1,\ldots,q\}\,,\nonumber\\
k&=&1,\ldots,p+1;\quad l=1,\ldots,q\,,
\end{eqnarray}
\begin{eqnarray}\label{MPOL2}
N^{(i)}_{k,q+1}\,&:=&\,N^{(i-1)}_{k,q+1}\,+\,{\rm med}
\{N^{(i-1)}_{kl}\,|\,l=1,\ldots,q\}\,,\nonumber\\
k&=&1,\ldots,p+1\,,
\end{eqnarray}
and define for $i=2,4,6,\ldots,$ 
\begin{eqnarray}\label{MPOL3}
N^{(i)}_{kl}\,&:=&\,N^{(i-1)}_{kl}\,-
\,{\rm med}\{N^{(i-1)}_{kl}\,|\,k=1,\ldots,p\}\,,\nonumber\\
k&=&1,\ldots,p;\quad l=1,\ldots,q+1\,,
\end{eqnarray}
\begin{eqnarray}\label{MPOL4}
N^{(i)}_{p+1,l}\,&:=&\,N^{(i-1)}_{p+1,l}\,+\,{\rm med}
\{N^{(i-1)}_{kl}\,|\,k=1,\ldots,p\}\,,\nonumber\\
l&=&1,\ldots,q+1\,.
\end{eqnarray}
Here, ${\rm med}\{x_1,\ldots,x_n\}$ is the median of
$\{x_1,\ldots,x_n\}$. The algorithm starts with 
\begin{equation}\label{MPOLA}
N^{(0)}_{kl}\,=\,\left\{ \begin{array}{r@{\quad:\quad}l}
N_{kl} & k=1,\ldots,p;\quad l=1,\ldots,q \\ 0 & {\rm elsewhere.}
\end{array}\right.
\end{equation}
Usually the method converges after three to five iterations.

\begin{figure}
\psfig{figure=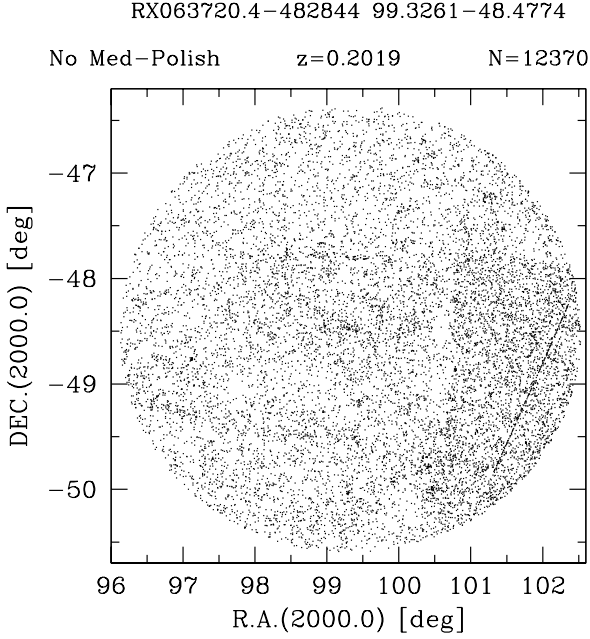,height=9cm}
\caption{{\small Angular distribution of the COSMOS galaxies with 
$b_J\le 20.5$\,mag centered around the ROSAT X-ray source
RX063720.4-482844. The data are collected from four different Schmidt
plates. Multiple detections of galaxies in the overlaps of adjacing
plates are rejected. The redshift of the central cluster of galaxies
is $z=0.2019$ measured in the course of an ESO Key Programme (Guzzo et
al. 1995).}}\label{F_NOMED}
\end{figure}

\begin{figure}
\psfig{figure=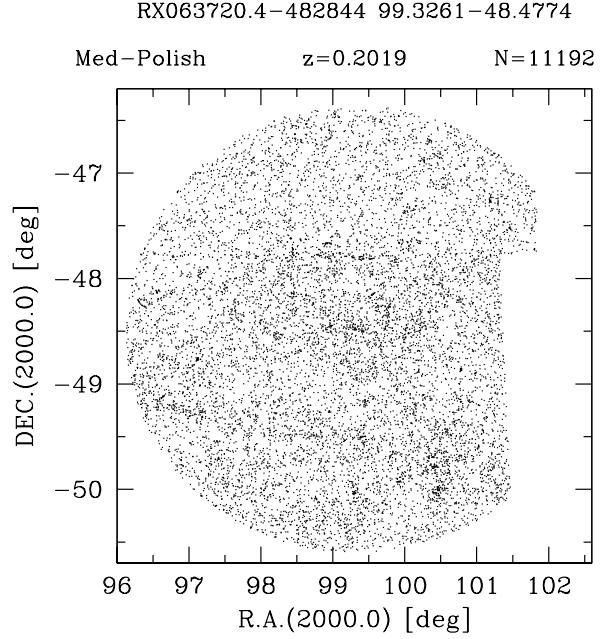,height=9cm}
\caption{{\small As Fig.\,\ref{F_NOMED} after median polishing. The 
weak indication of a smaller surface number density in the north
compared to the south might be attributed to a small error in the
magnitude zero-point. Median polishing was adjusted to accept such
large-scale deviation from stationarity which are not uncommon in the
COSMOS data base. Note that the excised region is smaller than the
dark region in Fig.\,\ref{F_NOMED} due to the plate
overlap.}}\label{F_MED}
\end{figure}

After median polishing the original $p\times q$ data matrix is
replaced by a $p\times q$ residual matrix, $\Delta_{kl}$, and by the
$p+q+1$ extra elements containing the deviations along the rows,
$\{r_k\}$, the deviations along the columns, $\{c_l\}$, and the
average background level, $\{a\}$. These quantities are related by
\begin{eqnarray}\label{MPOL4}
N_{kl}\,=\,a\,+\,r_k\,+\,c_l\,+\Delta_{kl}\,,\nonumber\\
k=1,\ldots,p;\quad l=1,\ldots,q\,.
\end{eqnarray}
Equation\,(\ref{MPOL4}) shows that $a$ describes the overall median
surface number density of galaxies in the area covered by the spatial
indices $k$ and $l$. Correspondingly, the projected median surface
number density profile in excess to the global median $a$ along, e.g.,
the Right Ascension and the Declination axes, is thus given by the
$r_k$ and by the $c_l$ vector, respectively. The residual matrix
$\Delta_{kl}$ gives the remaining deviations from (first-order)
stationarity. The three effects are unbiased estimates if the
coordinates of the individual galaxies are transformed into equal-area
Hammer Aitoff coordinates.

Figures\,\ref{F_NOMED} and \ref{F_MED} illustrate the performance of
the filter. In this example the COSMOS galaxy data of four adjacing
Schmidt plates (\ref{F_NOMED}) are combined. Median polishing detects
the artifical inhomogeneity at ${\rm R.A.}>100.5$ degrees, rejects the
corresponding data from further analyses, and starts again with a new
combination of the COSMOS data from the remaining Schmidt plates
(\ref{F_MED}). The algorithm can be regarded as a standard procedure
in spatial data analyses and is thus well studied both mathematically
and in many practical applications (see the examples and references
collected by Cressi 1993).

\end{appendix}
                                                                                
\end{document}